\documentclass[twocolumn]{aastex63}
\usepackage{wasysym}
\usepackage{colortbl}

\usepackage{array}

\definecolor{mgm}{rgb}{0.0, 0.0, 1.0}

\definecolor{sd}{rgb}{0.5, 0.0, 1.0}

\definecolor{anf}{rgb}{0.5, 0.5, 0.5}

\newcommand\degree{^{\circ}}
\newcommand\name{\texttt{exoMMR}}

\begin{document}

\shorttitle{exoMMR}
\shortauthors{MacDonald et al. }

\correspondingauthor{Mariah G. MacDonald}
\email{macdonam@tcnj.edu}

\title{\name: a New Python Package to Confirm and Characterize Mean Motion Resonances}
\author[0000-0003-2372-1364]{Mariah G. MacDonald}
\affiliation{Department of Astronomy \& Astrophysics, Center for Exoplanets and Habitable Worlds, The Pennsylvania State University, University Park, PA 16802, USA}
\affiliation{Department of Physics, The College of New Jersey, 2000 Pennington Road, Ewing, NJ 08628, USA}
\author[0009-0001-2321-7865]{Michael S. Polania Vivas}
\affiliation{Department of Physics, The College of New Jersey, 2000 Pennington Road, Ewing, NJ 08628, USA}
\author[0000-0001-5592-6220]{Skylar D'Angiolillo}
\affiliation{Department of Physics, The College of New Jersey, 2000 Pennington Road, Ewing, NJ 08628, USA}
\author[0009-0003-6956-4066]{Ashley N. Fernandez}
\affiliation{Department of Physics, The College of New Jersey, 2000 Pennington Road, Ewing, NJ 08628, USA}
\author[0000-0002-8974-8095]{Tyler Quinn}
\affiliation{Department of Astronomy \& Astrophysics, Center for Exoplanets and Habitable Worlds, The Pennsylvania State University, University Park, PA 16802, USA}

\begin{abstract}
The study of orbital resonances allows for the constraint of planetary properties of compact systems. We can predict a system's resonances by observing the orbital periods of the planets, as planets in or near mean motion resonance have period ratios that reduce to a ratio of small numbers. However, a period ratio near commensurability does not guarantee a resonance; we must study the system's dynamics and resonant angles to confirm resonance. Because resonances require in-depth study to confirm, and because two-body resonances require a measurement of the eccentricity vector which is quite challenging, very few resonant pairs or chains have been confirmed. We thus remain in the era of small number statistics, not yet able to perform large population synthesis or informatics studies. To address this problem, we build a python package to find, confirm, and analyze mean motion resonances, primarily through N-body simulations. We then analyze all near-resonant planets in the Kepler/K2 and TESS catalogues, confirming over 60 new resonant pairs and various new resonant chains. We additionally demonstrate the package’s functionality and potential by characterizing the mass-eccentricity degeneracy of Kepler-80g, exploring the likelihood of an exterior giant planet in Kepler-80, and constraining the masses of planets in Kepler-305. We find that our methods overestimate the libration amplitudes of the resonant angles and struggle to confirm resonances in systems with more than three planets. We identify various systems that are likely resonant chains but that we are unable to confirm, and highlight next steps for exoplanetary resonances.

\end{abstract}


\section{Introduction} \label{sec:intro}

Two planets are in mean motion resonance (MMR) with one another when they repeatedly conjunct at the same place, allowing them to exchange both energy and angular momentum. MMRs allow otherwise unstable configurations to persist and act as a potential well, resisting change from small perturbations. 

We define the critical resonant angle for two bodies as:
\begin{equation}\label{twobodyres}
\Theta_{b,c} = j_1\lambda_b + j_2\lambda_c + j_3\omega_b + j_4\omega_c + j_5\Omega_b + j_6\Omega_c
\end{equation}
\noindent where $\lambda_p$ is the mean longitude of planet $p$, $\omega_p$ is the argument of periapsis, $\Omega_p$ is the longitude of the ascending node, $j_i$ are coefficients which sum to zero, and planet $b$ orbits interior to planet $c$.

If a system contains more than two planets in resonance, the planets can be in a resonant chain, either a chain of two-body resonances or in a three-body resonance. The zeroth-order three-body resonance can be defined as the difference between two consecutive two-body resonances:
\begin{equation}\label{threebodyres}
    \phi_{b,c,d} = \Theta_{c,d} - \Theta_{b,c} = m\lambda_d - (m + n)\lambda_c + n\lambda_b
\end{equation}
\noindent where $\lambda_p$ is the mean longitude of planet $p$, and $m$ and $n$ are integers. 

For planets in resonance, the two-body and/or three-body resonant angle will librate about some center with some amplitude. From this libration amplitude, we can learn additional information about the system's formation history. Small libration amplitudes indicate low energy of the resonance and overall a close proximity to exact resonance, achieved by smooth and dissipative formation \citep{hadden2020}, whereas large libration amplitudes could be a consequence of perturbations from an additional planet \citep[e.g., ][]{dawson2021}, overstable librations \citep{goldreich2014}, or stochastic forcing \citep[e.g., ][]{rein2009}.

If two planets are in resonance with one another, such a configuration requires a specific parameter space. Because of this, we are able to constrain planetary masses and orbits to those that allow for resonance \citep{macdonald2022}.

The two bodies in resonance will have orbital periods whose ratio reduces to a ratio of small integers, providing a straightforward method for identifying potential resonances; however, the two planets need not be at exact commensurability, nor does exact commensurability guarantee resonance, as the resonant configuration depends on other factors as well, such as the planets' masses and eccentricities. Because of this complexity, we cannot simply assume that any two adjacent planets near commensurability are resonant and instead must study their dynamics to confirm a resonance. Such a study is tedious and sometimes not possible, since the eccentricity vector is challenging to constrain, and so most systems remain classified as ``near-resonant.'' 

Traditionally, mean motion resonance in exoplanets is confirmed by integrating forward the solutions to the system's radial velocities \citep[RV, e.g., ][]{nelson2016} or transit timing variations \citep[TTVs, e.g., ][]{macdonald2016}. Such a process, however, requires that the system has detectable RVs or TTVs and that the signal from the planet-planet perturbations is sufficiently large to favor non-Keplerian orbits. One additional method of confirming resonance is modeling all possible solutions to the system \citep{macdonald2022, quinn2023}. Although computationally intensive, this brute-force method can confirm resonance if all solutions lead to resonance.

Following the methods of \citet{macdonald2022} and \citet{quinn2023}, we create a python package to identify, confirm, and characterize additional and known resonances in exoplanetary systems that we call \name~\citep{exommr}.

In Section~\ref{sec:code}, we discuss the structure and performance of \name, and we provide numerous examples of verification and utility of the software in Section~\ref{sec:examples}, including identifying new resonant systems. We discuss the limitations of this software and our methods in Section~\ref{sec:limitations} and summarize and conclude in Section~\ref{sec:conclusion}.

\section{Overview of the code structure}\label{sec:code}

\name~performs various functions associated with finding, confirming, and characterizing mean motion resonances in exoplanetary systems. Each of these functions, summarized below, can be performed separately from one another.

\subsection{Identifying resonance}
We can quantify a planet pair's proximity to mean motion resonance, sometimes referred to as the resonance offset, by taking the difference between the observed motions and the mean-motion commensurability:

\begin{equation}
    \Delta_{(p+q)/p} = \frac{n_i}{n_{(i+1)}} - \frac{(p+q)}{p}
\end{equation}
where $p$ and $q$ describe the resonance, $n$ is the mean motion, and $i$ is the planet closer to the primary. Typically, a proximity of 0.01 or less is associated with a resonant pair, although this proximity can be shifted by tidal dissipation and perturbations from an additional resonant pair \citep[e.g., ][]{macdonald2016}.

In practice, \name~requires an array of orbital periods to calculate the proximity to resonance; it will test all first- and second-order two-body mean motion resonances, returning the resonant angle and the pair's proximity to that resonance.

\subsection{Creating suite of N-body simulations}
Most of the functions of \name~require numerous models of the system. Although this requirement can be satisfied with posteriors from radial velocity, transit timing variations, or photodynamical fitting, it can also be met with a suite of \textit{N}-body simulations. \name~will create and run a suite of rebound simulations \citep{rebound}, pulling planetary, stellar, and simulation parameters from an input file. The software is structured to run the suite of simulations as a SLURM job array, but the jobs can be run in series or with another resource manager.

\name~will default to the following options: 
Stellar masses will be fixed to the value provided. Planet masses, orbital periods, eccentricities, and inclinations will be drawn from independent, normal distributions centered on the nominal values with standard deviations equal to the uncertainties. Each planet is initialized with a longitude of the ascending node drawn from a uniform distribution U[0,$\pi$] and a mean longitude calculated from the given transit epoch or mid-transit time associated with the orbital period fit. Each suite of simulations will consist of 500 simulations, integrated for 1e6 years, with the WHFast integrator \citep{whfast} and an integration timestep of 5\% the smallest orbital period. 

\newpage

\subsection{Confirming resonance and resonant chains}
Given a rebound simulation, \name~will calculate the center of a resonant angle as the median and the amplitude of the libration as twice the standard deviation of the angle over a few years. The angles will be wrapped between [0$\degree$, 360$\degree$] and between [-180$\degree$, 180$\degree$], and the angle with the smallest calculated amplitude will be taken. A simulation is marked as resonant if the amplitude of libration is less than 150$\degree$.

\name~can then calculate the statistics of the angle across all simulations, including the percentage of simulations with that angle librating, then the median and uncertainties of the libration center and amplitude. Once the individual angles are characterized, \name~is able to study the possibility of resonant chains; here, we define a resonant chain as either two or more consecutive librating two-body angles or three-body angles. \name~will return the percentage of simulations that result in a three-body, four-body, etc., resonant chain, as well as the percentage of simulations where each planet is dynamically decoupled. 

For this work, we confirm resonance if 90\% or more of the simulations result in librating angles; although this number is fairly arbitrary, we caution against reducing it. {We discuss the potential limitations of this cut-off more in Section~\ref{sec:cutoff}.}

\subsection{Constraining parameters with resonance}

Using statistical tests, we can confirm whether there is a significant difference between solutions that lead to resonance and those that do not. Following \citet{macdonald2021} and \citet{macdonald2022}, we compare two samples of a parameter, split by whether or not a resonant angle is librating, using both a Kolmogorov–Smirnov test and an Anderson-Darling test. Both of these tests explore the null hypothesis that the two samples are drawn from the same population, so a resulting p-value less than $\alpha=5\%$ allows us to reject this hypothesis.

\subsection{Exploring chain formation}
Resonant chains are often seen as the hallmarks of convergent migration, as planets will migrate until they lock into resonances then the resonant pair will migrate together, locking in additional planets \citep[][]{cossou2013}. However, resonant chains do not require such long-distance migration; dissipation from a disk, tides, or planet-planet scattering can damp a planet's eccentricity and cause slight migration, also resulting in chains of resonances \citep{macdonald2016,dong2016, macdonald2018}.

Following the methods outlined in \citet{macdonald2018}, \name~creates and runs suites of N-body simulations. Each simulation initializes the planets out of resonances and then damps the semi-major axes and eccentricities of the planets, following the prescription in \citet{papaloizou2000} and using the \texttt{modify\_orbits\_forces} implementation \citep{kostov2016} in {\textsc \tt REBOUNDx 3.1.0} \citep{tamayo2020}; for the migration simulations, these forces are applied only to the outermost planet under the assumption that it is a shorter migration timescale than the other planets in the system. For simulations with only eccentricity damping, the damping is applied to each planet. The migration and eccentricity damping timescales are drawn from independent log-normal distributions whose bounds are user-defined.

\section{Test problems and utility}\label{sec:examples}
\subsection{Recovering known resonances}\label{sec:known}
To verify the usability of \name, we study two well-studied resonant chain systems: Kepler-223 and Kepler-80. For each system, we run 500 N-body simulations, drawing the planet masses and orbital elements from independent normal distributions, centered around the nominal values with widths of the uncertainties constrained by \citet{mills2016} and \citet{macdonald2016}. We integrate at 5\% the innermost planet's orbital period using the WHFast integrator \citep{whfast}. 

After 1~Myr, we study the two- and three-body resonant angles that correspond to the orbital period commensurabilities. We estimate the angle amplitude as twice the standard deviation of the angle over a period of 20 years\footnote{2$\sigma$ resulted in an amplitude that was least biased by long-term linear changes to the libration center and by cycling in and out of resonance}, and constrain the number of simulations in which each angle is librating. We summarize our results in Table~\ref{tab:known}.

\begin{deluxetable}{lccc}
\renewcommand{\arraystretch}{1.0}
\tablecolumns{4}
\tabletypesize{\footnotesize}
\tablecaption{ Known Resonant Systems \label{tab:known}}
\tablehead{
\colhead{System} &
\colhead{\% librating} &
\colhead{Center} &
\colhead{Amplitude} 
}
\startdata
Kepler-80 & & & \\
$\Theta_{1,2}$	&	85.00\%	& 	-0.06	$_{-	0.42	}^{+	0.61	}$ &	94.66	$_{-	44.32	}^{+	38.63	}$ \\
$\Theta_{2,3}$	&	40.00\%	&	-0.95	$_{-	6.77	}^{+	3.66	}$ &	131.26	$_{-	10.54	}^{+	33.38	}$ \\
$\Theta_{3,4}$	&	63.00\%	&	-0.81	$_{-	8.46	}^{+	5.97	}$ &	120.50	$_{-	20.59	}^{+	20.29	}$ \\
$\phi_1$	&	16.00\%	&	176.38	$_{-	5.95	}^{+	7.91	}$ &	126.03	$_{-	15.51	}^{+	37.52	}$ \\
$\phi_2$	&	30.00\%	&	56.07	$_{-	31.10	}^{+	132.47	}$ &	100.78	$_{-	39.21	}^{+	23.81	}$ \\
\hline
Kepler-223 & & & \\
$\Theta_{1,2}$	&	39.81\%	&	-0.04	$_{-	24.26	}^{+	26.87	}$ &	122.85	$_{-	19.73	}^{+	30.27	}$ \\
$\Theta_{2,3}$	&	31.52\%	&	0.51	$_{-	24.73	}^{+	27.61	}$ &	127.39	$_{-	17.75	}^{+	28.02	}$ \\
$\Theta_{3,4}$	&	29.62\%	&	-0.81	$_{-	40.31	}^{+	24.68	}$ &	128.80	$_{-	12.93	}^{+	22.70	}$ \\
$\phi_1$	&	69.19\%	&	52.90	$_{-	130.27	}^{+	87.42	}$ &	84.48	$_{-	43.69	}^{+	23.46	}$ \\
$\phi_2$	&	32.94\%	&	67.49	$_{-	102.48	}^{+	126.51	}$ &	116.13	$_{-	25.73	}^{+	34.01	}$ 
\enddata
\tablecomments{The results of \name~for two known resonant chain systems. We should the percentage of simulations where each angle is librating along with the center and amplitude of the libration.}
\end{deluxetable}

Both Kepler-223 and Kepler-80 have well-studied and confirmed four-body resonant chains. However, \name~is unable to confirm such a chain. Although each two-body angle librates in a significant percentage of simulations,  we are not able to confirm any of these angles since no angle librates in more than 90\% of the simulations. In addition, the three-body resonant angles librate in a fair fraction of simulations for both systems, but no angle librates in a large enough fraction to consider the system in resonance.

We discuss the implications of our inability to recover these resonant chains in Section~\ref{sec:limitations}.

\subsection{Systems without resonance}\label{sec:known-not}
In addition to studying systems with known resonances, we validate the effectiveness of \name~by studying Kepler-11. Kepler-11 is a G-type star hosting six super-Earths. While five of these planets all orbit their star within 50 days, making this system one of the first compact and dynamically cold systems discovered \citep{lissauer2011-k11}, the planets in this system are well-studied and confirmed to \textit{not} be in resonance with one another \citep{lissauer2011-k11, migaszewski2012,mahajan2014}. We run 500 \textit{N}-body simulations of the system and its five inner planets. For each planet, we draw its mass and orbital elements from independent normal distributions centered on the nominal values and with widths equal to the uncertainties in \citet{lissauer2013}. We integrate for 1~Myr with a timestep of 5\% the innermost planet's orbital period using the WHFast integrator \citep{whfast}. We assume a stellar mass of 1.04$M_{\odot}$ \citep{stassun2019}.

Of the 500 simulations, \name~marks six simulations (1.2\%) as containing resonances. In each of these simulations, the two-body angle $\Theta_{c,d}= 5\lambda_d - 4\lambda_c - \omega_c$ and the three-body angle $\phi_{d,e,f} = 4\lambda_f-7\lambda_e+3\lambda_d$ librate with large amplitudes of $76.74^{+32.38}_{-32.38}$ and $69.23^{+2.44}_{-2.44}$, respectively. Although we only have these six simulations, we find no statistical evidence of preferred masses or orbits that lead to this resonance and instead find it likely that these resonant angles are switching between librating and circulating. 

\vspace{3mm}

\subsection{Constraining outer companions in Kepler-80}

Kepler-80 is a K-dwarf that hosts six known transiting exoplanets, with orbital periods ranging between 1.0 and 14.7~days and radii between 1.2 and 2.2~$R_{\oplus}$ \citep{macdonald2016}. Four of these planets are locked in a chain of three-body mean motion resonances, and the outermost planet is likely also in resonance \citep{shallue2018, macdonald2021}. The two-body angles associated with the commensurabilities in this system do not librate \citep[][]{macdonald2016, macdonald2021, weisserman2023}, making Kepler-80 a relatively unique system and a useful test ground for planetary formation and evolution. 

Since the resonances in this system are well-studied, we are able to leverage the dynamics to constrain any undetected outer companions. We model the five outermost planets and an injected theoretical planet. We draw the orbital period of this injected from a uniform distribution spanning 15 to 60 days, the mass from 0.5$M_{\oplus}$--5$M_J$, and initialize the planet with a dynamically cold orbit ($e = 0.0$, $i=90\degree$). We draw the masses and orbital elements of the known planets from independent normal distributions as described in Section~\ref{sec:code} and integrate for $5\times10^5$ years at a timestep of 5\% the inner planet's orbital period. 

We constrain the feasibility of the injected planet with two criteria, the system's stability and the known resonance. If the new planet results in orbital evolution and a close encounter or ejection, it could not possibly exist. Similarly, if the new planet disrupts or breaks the known three-body resonances and causes them to circulate, it could not exist. We therefore restrict the ranges of possible masses and orbital resonances with these criteria.

\begin{figure*}
    \centering
    \includegraphics[width=0.49\textwidth]{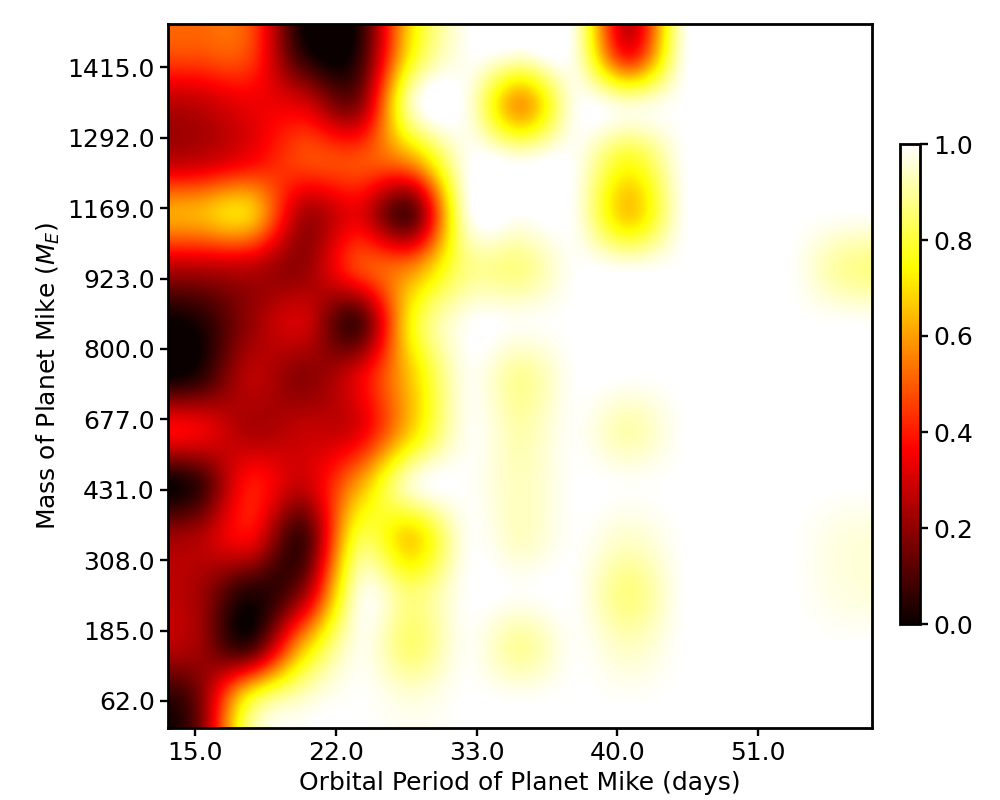}
    \includegraphics[width=0.49\textwidth]{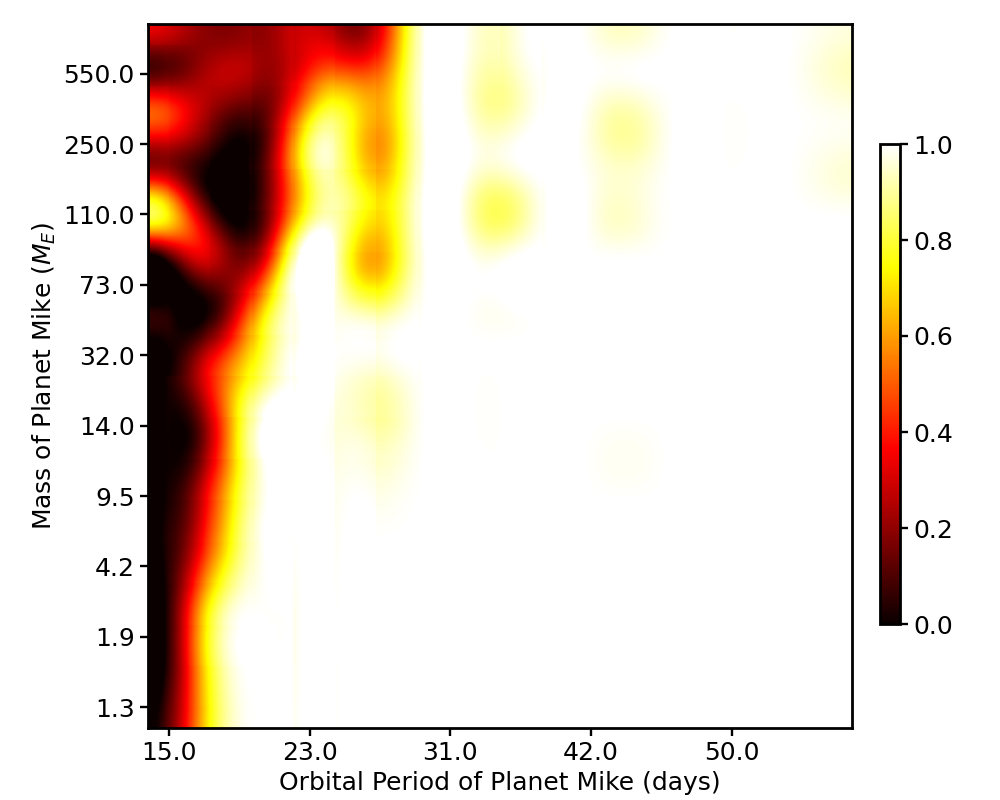}
    \caption{Left) Probability of an injected planet in Kepler-80. Here, we color the region based on the percentage of simulations that remained stable and have librating three-body resonant angles at the end of the $5\times10^5$ year integration. We explore a range of masses from $0.5M_{\oplus}$--$5M_J$ and initialize the injected planet at $e=0.0$ and $i=90\degree$. We find that any planet with $P>50$ days, regardless of mass, is allowed. Planets closer to this must be part of the resonant chain if they are massive; otherwise the resonances are broken and the system often becomes unstable. Interior to $P<20$ days, the system usually goes unstable from an excited planet. }
    \label{fig:K80-mike-hm}
\end{figure*}

We summarize our results as heat maps in Figure~\ref{fig:K80-mike-hm}. We find that any planet close to Kepler-80g ($P<20$ days), regardless of mass, is likely to cause instability. We also find that any planet exterior to $P>50$ days would not disrupt the resonances or cause system instability. Curiously, a massive planet ($M_p>0.25M_J$) with $20<P<50$ would need to participate in the resonant chain to avoid instability or breaking the existing resonances.

\subsection{Exploring the mass-eccentricity degeneracy of Kepler-80g}
Discovered via neural nets by \citep{shallue2018}, Kepler-80g is the outermost known planet orbiting its K-type host. Due to its low signal-to-noise ratio of 8.6 \citep{shallue2018}, the planet's orbital period and radius are constrained to a much lower precision than is typical for transiting planets, with $P=14.65\pm0.001$~days and $R_p=1.05^{+0.22}_{-0.24}~R_{\oplus}$. 

Kepler-80g's orbital period suggests that it likely continues the chain of MMRs seen in the rest of the system. To confirm this resonance and further characterize the planet, \citet{macdonald2021} photodynamically fit the system. They recover a radius of $R_p=1.05^{+0.22}_{-0.24}~R_{\oplus}$, an orbital period of $P=14.65\pm0.001$ days, and a mass of $M_p=0.065^{+0.044}_{-0.038}~M_{\oplus}$. This mass in conjunction with the radius estimate suggests a low density planet that is atypical of terrestrial-size planets. Combined with the high precision of the mass estimate, it is likely that this planet was overfit. In addition, \citet{macdonald2021} find an eccentricity of $e=0.13$, significantly greater than the eccentricities of the other planets in the systems and greater than most resonant, small planets. We find it likely, then, that \citet{macdonald2021} report a mass that is far too low. We aim to constrain the possible ranges of mass and eccentricity that this planet must need to exist, as well as to not disrupt the resonance of its neighboring planets within Kepler-80.

To break this mass-eccentricity degeneracy, we run a total of 1200 \textit{N}-body simulations. We draw Kepler-80g's mass and eccentricity from independent uniform distributions of U[0.0,1.0]~$M_{\oplus}$ and U[0.0,0.1], respectively, draw its orbital period from a Gaussian distribution of N[14.651,0.001] days, and initialize its inclination at 88.26$\degree$. We draw the masses and orbital parameters for the other planets in the system from independent normal distributions centered around the nominal values from \citet{macdonald2021} and fix the stellar mass to 0.73~$M_{\odot}$ \citep{macdonald2016}. 
We integrate for 1~Myr with a timestep of 5\% the inner planet's orbital period using the WHFast integrator \citep{whfast}. We stop integrating if any planet experiences a close encounter or if any planet's eccentricity exceeds 0.9. We then analyze each simulation for resonance, looking for libration of the two three-body resonant angles $\phi_1=3\lambda_b-5\lambda_e+2\lambda_d$ and $\phi_2=2\lambda_c-3\lambda_b+\lambda_e$. 

\begin{figure*}
    \centering
    \includegraphics[width=0.47\textwidth]{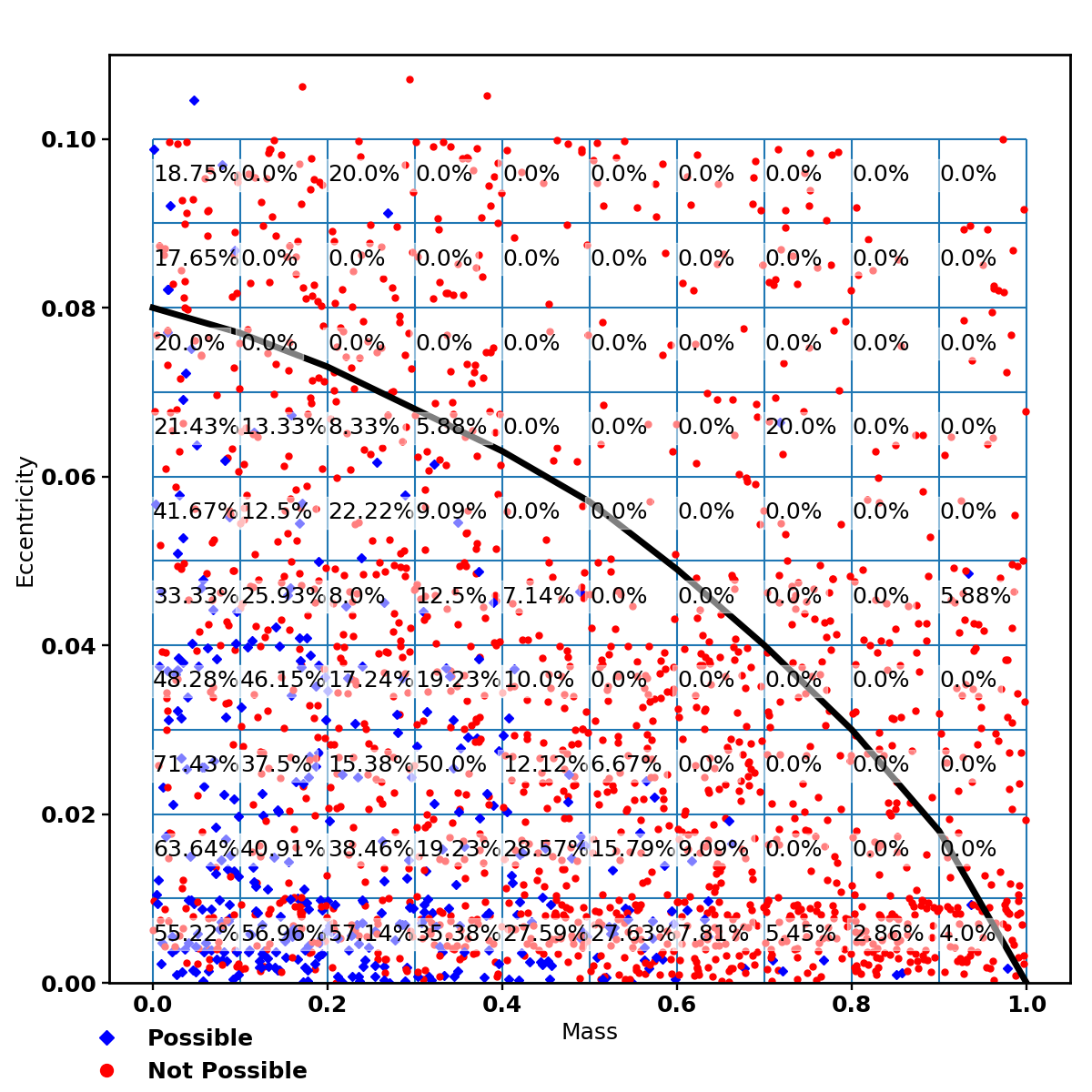}
    \includegraphics[width=0.47\textwidth]{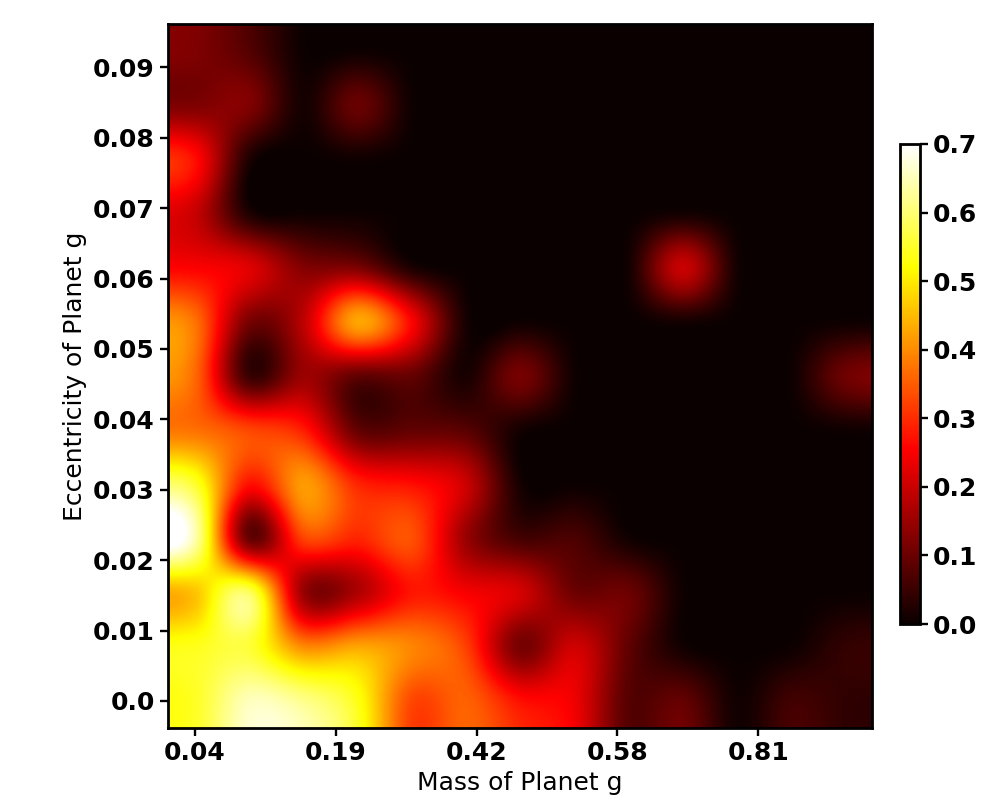}
\caption{Results of our simulations of Kepler-80g. Here, we draw the mass and eccentricity from independent uniform distributions of U[0.0,1.0]~$M_{\oplus}$ and U[0.0,0.1], respectively. Left) Entire parameter space, marking simulations as ``Not Possible'' (red) if the simulation went unstable or if the known three-body resonances circulated and ``Possible'' (blue) if both criteria are met. Right) Heat map of same results, where black shows 0\% of simulations met both stability and resonant criteria and white shows 100\% of simulations met both criteria. We find that, for Kepler-80g to be $M_p>0.5~M_{\oplus}$, the eccentricity must be small, $e<0.005.$.}
    \label{fig:K80g}
\end{figure*}

We show our results in Figure~\ref{fig:K80g}. We find that Kepler-80g must be relatively low mass with low eccentricity for the system to remain stable with its resonances intact. If the planet is more massive than 0.5~$M_{\oplus}$, corresponding to a minimum bulk density of $\rho=2.38~g/cm^3$ with an assumed radius of $R_p=1.05~R_{\oplus}$, we find the eccentricity must be small, $e<0.005.$ Eccentricities larger than this result in a disruption of the known three-body resonances, causing one or both angle to circulate instead of librate. Specifically, we constrain the mass and eccentricity to $0.20^{+0.25}_{-0.14}~M_{\oplus}$ and $0.01^{+0.03}_{-0.007}$, respectively.

The results for mass and eccentricity that we derive from dynamics are still far from realistic. Assuming a planetary radius of $R_p=1.05~R_{\oplus}$, Kepler-80g would have a bulk density of $0.927^{+1.19}_{-0.65}~g/cm^3$, significantly less dense than most planets. We do, however, recover an eccentricity for Kepler-80g that is much smaller than the estimate from \citet{macdonald2021} of $e=0.13$ and is more inline with other compact systems.

Recently, \citet{weisserman2023} re-visited the Kepler-80 system, performing an analysis similar to \citet{macdonald2016} but including Kepler-80g in their fits. When they allow the eccentricity vectors of the five planets to float, they recover a mass of $0.8^{+0.8}_{-0.6}~M_{\oplus}$ and an eccentricity of $0.02^{+0.03}_{-0.02}$. This eccentricity is consistent with our dynamically-derived estimate of $0.01^{+0.03}_{-0.007}$, suggesting this to likely be accurate. This mass estimate is larger than our mass estimate, although consistent within 1$\sigma$, suggesting that we are, still, underestimating the mass.

We have ultimately constrained the mass and eccentricity of Kepler-80g using the system's dynamics. We use this example as a proof-of-concept that other less-studied systems can have their parameters dynamically constrained in the absence of detectable TTVs or RVs. 

\subsection{Confirming new resonances}\label{sec:findnew}

We explore all Kepler, K2, and TOI systems for mean motion resonances.
For each consecutive planet pair, we calculate the proximity to resonance, using the periods reported in the \texttt{Exoplanet Archive} \citep{PSCompPars, k2pandc, koidr25}. We then study systems with at least one planet pair that is wide of a resonance with a proximity to resonance less than 0.2\footnote{We do not study systems if the only near-resonant pair is inside the resonance as these are unlikely to be resonant.} We then down-select this target list, removing systems with known resonances. 

For each system, we run a suite of 500 \textit{N}-body simulations. We assume a stellar mass from that reported in each catalog and draw the planetary parameters from independent normal distributions that are centered on the nominal values reported in the respective reference. For planets without mass estimates, we use the mass-radius relationship from \cite{weiss2014}\footnote{Although resonant state indeed depends on planetary mass, the uncertainty in mass that results from assuming a mass-radius relationship is typically smaller than the resonance width for small planets; therefore, our results are not sensitive to this mass-radius relationship for the majority of the planets included in this study.}. Otherwise, we use the mass estimates from the primary reference on the \texttt{exoplanet archive}. We summarize our starting conditions in Table~\ref{tab:conditions}.

We use the WHFast integrator with a timestep set to 5\% of the inner planet's orbital period and integrate for 1~Myr\footnote{{We select 1~Myr to help ensure long-term stability while reducing computational costs. We have found no statistically significant difference in results between 1.0 and 10~Myr \citep[e.g., ][]{quinn2023}.}} or until instability (close encounter). We then study each simulation for resonant behaviour; we look for libration of each resonant angle based on a libration amplitude that is less than 150$\degree$, and we confirm a resonance if the resonant angle librated in 90+\% of our \textit{N}-body simulations. 

Overall, we confirm 66 new resonances in 60 systems. We summarize these new resonances in Table~\ref{tab:confirm}. For completeness, we summarize resonances that we explored but cannot confirm in the Appendix in Table~\ref{tab:nonconfirm}\footnote{{Although we classify systems as resonant, we caution against labeling the other systems as ``nonresonant.'' Instead, we say we are unable to confirm resonance and mark some as potentially resonant.}}.

\startlongtable
\begin{deluxetable*}{lcccccc}
\renewcommand{\arraystretch}{1.0}
\tablecolumns{7}
\tablewidth{\textwidth}
\tabletypesize{\footnotesize}
\tablecaption{New Mean Motion Resonances \label{tab:confirm}}
\tablehead{
\colhead{System} &
\colhead{Planets} &
\colhead{Resonance} &
\colhead{\% librating} &
\colhead{Center} &
\colhead{Amplitude} &
\colhead{Notes} 
}
\startdata
HD 28109	&	02, c	&	3:2	&	100.0	&	-1.63	$^{+	17.97	}_{-	14.35	}$	&	120.83	$^{+	5.95	}_{-	10.85	}$	&	2, 4	\\
HIP 41378	&	b, c	&	2:1	&	100.0	&	-0.012	$^{+	0.42	}_{-	0.39	}$	&	69.61	$^{+	28.78	}_{-	22.08	}$	&	3, 4	\\
HD 191939	&	c, d	&	4:3	&	100.0	&	0.21	$^{+	5.26	}_{-	5.73	}$	&	114.23	$^{+	5.00	}_{-	3.86	}$	&	3	\\
HD 260655	&	b, c	&	2:1	&	100.0	&	0.047	$^{+	0.56	}_{-	0.57	}$	&	96.67	$^{+	8.90	}_{-	8.71	}$	&		\\
K2-80	&	b, d	&	3:2	&	100.0	&	0.29	$^{+	3.57	}_{-	3.58	}$	&	118	$^{+	6.20	}_{-	7.47	}$	&		\\
K2-178	&	02, b	&	2:1	&	91.2	&	0.011	$^{+	0.42	}_{-	0.45	}$	&	73.4	$^{+	47.97	}_{-	29.87	}$	&	2	\\
	&	b, 03	&	3:2	&	99.4	&	0.085	$^{+	4.13	}_{-	4.6	}$	&	117.96	$^{+	11.20	}_{-	12.65	}$	&	2	\\
K2-239	&	b, c	&	3:2	&	93.6	&	179.98	$^{+	0.767	}_{-	0.75	}$	&	99.89	$^{+	32.18	}_{-	48.06	}$	&		\\
K2-268	&	e, c	&	3:2	&	91.6	&	-0.072	$^{+	0.94	}_{-	0.89	}$	&	105.18	$^{+	32.03	}_{-	37.14	}$	&	1	\\
K2-285	&	b, c	&	2:1	&	100.0	&	0.0083	$^{+	0.29	}_{-	0.3	}$	&	71.89	$^{+	30.17	}_{-	27.00	}$	&		\\
Kepler-18	&	c, d	&	2:1	&	100.0	&	180.02	$^{+	1.00	}_{-	0.99	}$	&	79.56	$^{+	16.24	}_{-	21.13	}$	&	1, 3	\\
Kepler-23	&	b, c	&	3:2	&	97.2	&	-0.03	$^{+	2.06	}_{-	1.97	}$	&	132.43	$^{+	8.94	}_{-	8.11	}$	&	3	\\
Kepler-31	&	c, d	&	2:1	&	99.8	&	0.058	$^{+	2.06	}_{-	2.09	}$	&	81.5	$^{+	31.67	}_{-	23.63	}$	&	1, 3	\\
Kepler-32	&	e, b	&	2:1	&	99.8	&	-0.043	$^{+	0.63	}_{-	0.53	}$	&	77.55	$^{+	29.59	}_{-	18.81	}$	&	3	\\
	&	b, c	&	3:2	&	98.6	&	180.04	$^{+	0.69	}_{-	0.75	}$	&	129.56	$^{+	8.67	}_{-	14.08	}$	&		\\
Kepler-51	&	b, c	&	2:1	&	95.0	&	179.93	$^{+	2.41	}_{-	2.08	}$	&	82.38	$^{+	41.42	}_{-	35.52	}$	&	3	\\
	&	c, d	&	3:2	&	98.8	&	0.31	$^{+	14.53	}_{-	14.06	}$	&	122.03	$^{+	13.21	}_{-	10.82	}$	&		\\
Kepler-53	&	b, c	&	2:1	&	98.8	&	-0.060	$^{+	2.33	}_{-	2.34	}$	&	79.16	$^{+	33.32	}_{-	31.40	}$	&	1, 3	\\
Kepler-55	&	d, e	&	2:1	&	97.4	&	0.0078	$^{+	0.35	}_{-	0.40	}$	&	71.03	$^{+	41.92	}_{-	28.90	}$	&	3	\\
	&	b, c	&	3:2	&	100.0	&	0.96	$^{+	9.15	}_{-	11.02	}$	&	116.10	$^{+	7.77	}_{-	7.45	}$	&		\\
Kepler-62	&	e, f	&	2:1	&	100.0	&	0.045	$^{+	2.70	}_{-	2.76	}$	&	79.61	$^{+	29.49	}_{-	27.73	}$	&	1	\\
Kepler-83	&	b, c	&	2:1	&	99.8	&	0.035	$^{+	0.90	}_{-	0.91	}$	&	78.33	$^{+	37.96	}_{-	37.64	}$	&	1, 3	\\
Kepler-102	&	d, e	&	3:2	&	99.8	&	0.07	$^{+	1.87	}_{-	1.87	}$	&	89.23	$^{+	28.79	}_{-	27.22	}$	&	1, 3	\\
Kepler-104	&	b, c	&	2:1	&	99.8	&	-0.01	$^{+	0.57	}_{-	0.49	}$	&	69.59	$^{+	38.54	}_{-	35.80	}$	&	1, 3	\\
Kepler-105	&	c, 03	&	4:3	&	98.4	&	180.02	$^{+	2.03	}_{-	1.81	}$	&	123.16	$^{+	10.82	}_{-	17.93	}$	&	1, 2, 3	\\
Kepler-131	&	c, 03	&	3:2	&	93.8	&	-0.020	$^{+	0.61	}_{-	0.64	}$	&	112.43	$^{+	22.34	}_{-	26.83	}$	&	1, 2	\\
Kepler-138	&	b, c	&	4:3	&	100.0	&	0.24	$^{+	3.97	}_{-	3.85	}$	&	115.00	$^{+	5.98	}_{-	4.72	}$	&	3	\\
Kepler-154	&	f, d	&	2:1	&	99.2	&	0.040	$^{+	0.36	}_{-	0.40	}$	&	85.57	$^{+	31.14	}_{-	23.08	}$	&	1	\\
Kepler-169	&	c, d	&	4:3	&	99.6	&	-0.051	$^{+	1.36	}_{-	1.25	}$	&	95.63	$^{+	14.51	}_{-	19.83	}$	&	1	\\
Kepler-176	&	c, d	&	2:1	&	99.2	&	-0.45	$^{+	4.48	}_{-	4.13	}$	&	76.03	$^{+	37.57	}_{-	35.55	}$	&	1, 3	\\
Kepler-176	&	d, e	&	2:1	&	89.0	&	180.28	$^{+	9.36	}_{-	10.12	}$	&	92.76	$^{+	37.62	}_{-	41.72	}$	&	1	\\
Kepler-207	&	c, d	&	2:1	&	89.6	&	179.86	$^{+	3.35	}_{-	3.17	}$	&	86.56	$^{+	35.64	}_{-	28.32	}$	&	1	\\
Kepler-208	&	c, d	&	3:2	&	95.8	&	179.91	$^{+	1.89	}_{-	1.64	}$	&	112.74	$^{+	26.22	}_{-	31.69	}$	&		\\
Kepler-249	&	c, d	&	2:1	&	96.8	&	0.0078	$^{+	0.36	}_{-	0.37	}$	&	72.28	$^{+	45.87	}_{-	30.94	}$	&	1	\\
Kepler-254	&	c, d	&	3:2	&	99.2	&	-0.15	$^{+	2.41	}_{-	2.22	}$	&	116.40	$^{+	13.04	}_{-	14.18	}$	&	1, 3	\\
Kepler-305	&	b,c	&	3:2	&	99.6	&	0.011	$^{+	1.35	}_{-	1.24	}$	&	89.8	$^{+	21.33	}_{-	24.5	}$	&	1, 3	\\
Kepler-327	&	b, c	&	2:1	&	100.0	&	0.00074	$^{+	0.16	}_{-	0.18	}$	&	69.49	$^{+	32.38	}_{-	40.79	}$	&	3	\\
Kepler-332	&	b, c	&	2:1	&	99.2	&	0.007	$^{+	0.56	}_{-	0.58	}$	&	67.97	$^{+	25.10	}_{-	22.9	}$	&		\\
	&	c, d	&	2:1	&	95.4	&	-0.052	$^{+	4.48	}_{-	3.65	}$	&	69.34	$^{+	38.03	}_{-	20.83	}$	&		\\
Kepler-339	&	c, d	&	3:2	&	99.6	&	0.0047	$^{+	1.21	}_{-	1.29	}$	&	84.74	$^{+	18.27	}_{-	17.37	}$	&	3	\\
Kepler-341	&	b, c	&	3:2	&	95.4	&	0.017	$^{+	0.89	}_{-	0.99	}$	&	128.18	$^{+	10.34	}_{-	17.30	}$	&		\\
	&	c, d	&	3:2	&	99.8	&	-0.094	$^{+	1.09	}_{-	1.01	}$	&	129.75	$^{+	10.57	}_{-	19.71	}$	&		\\
Kepler-363	&	b, c	&	2:1	&	99.2	&	0.0029	$^{+	0.22	}_{-	0.24	}$	&	49.42	$^{+	33.52	}_{-	23.94	}$	&	1, 3	\\
Kepler-394	&	b, c	&	3:2	&	99.8	&	0.12	$^{+	0.83	}_{-	0.90	}$	&	89.87	$^{+	22.48	}_{-	32.39	}$	&		\\
Kepler-968	&	c,d	&	4:3	&	96.6	&	-0.04	$^{+	1.04	}_{-	1.01	}$	&	126.35	$^{+	10.07	}_{-	9.93	}$	&	1, 3	\\
Kepler-1518	&	b, 02	&	2:1	&	99.8	&	179.96	$^{+	3.15	}_{-	3.22	}$	&	51.65	$^{+	21.79	}_{-	22.48	}$	&	1	\\
Kepler-1581	&	02,04	&	3:2	&	90.4	&	0.044	$^{+	0.432	}_{-	0.515	}$	&	103.31	$^{+	27.34	}_{-	28.07	}$	&	2	\\
L 98-59	&	c, d	&	2:1	&	100.0	&	-0.028	$^{+	1.04	}_{-	0.95	}$	&	65.94	$^{+	24.45	}_{-	27.28	}$	&	4	\\
LHS 1678	&	c, 03	&	4:3	&	100.0	&	-0.0071	$^{+	0.67	}_{-	0.66	}$	&	119.29	$^{+	5.83	}_{-	5.89	}$	&	2	\\
TOI-178	&		&		&	97.2	&	51.9	$^{+	45.39	}_{-	147.1	}$	&	59.64	$^{+	29.95	}_{-	35.34	}$	&	1, known	\\
TOI-270	&	c, d	&	2:1	&	100.0	&	-0.14	$^{+	3.49	}_{-	3.12	}$	&	94.11	$^{+	7.44	}_{-	18.05	}$	&		\\
TOI-406	&	02, 01	&	2:1	&	100.0	&	180.06	$^{+	7.25	}_{-	7.45	}$	&	96.71	$^{+	5.28	}_{-	5.11	}$	&	2	\\
TOI-561	&	c, f**	&	3:2	&	100.0	&	0.067	$^{+	1.07	}_{-	1.17	}$	&	120.3	$^{+	9.11	}_{-	13.26	}$	&	3	\\
TOI-663	&	02, 03	&	3:2	&	98.0	&	0.045	$^{+	1.07	}_{-	1.17	}$	&	126.81	$^{+	11.42	}_{-	14.45	}$	&	2	\\
TOI-1097	&	01, 02	&	3:2	&	100.0	&	0.28	$^{+	1.88	}_{-	2.46	}$	&	118.40	$^{+	7.08	}_{-	10.94	}$	&	2	\\
TOI-1130	&	b, c	&	2:1	&	100.0	&	-0.027	$^{+	0.60	}_{-	0.62	}$	&	103.17	$^{+	4.99	}_{-	4.28	}$	&		\\
TOI-1246	&	d, e	&	2:1	&	100.0	&	0.17	$^{+	2.25	}_{-	2.38	}$	&	62.37	$^{+	24.83	}_{-	24.35	}$	&	1, 3	\\
TOI-1445	&	02, 01	&	2:1	&	100.0	&	0.015	$^{+	0.64	}_{-	0.68	}$	&	78.77	$^{+	19.69	}_{-	23.13	}$	&	2	\\
TOI-1453	&	02, 01	&	3:2	&	100.0	&	0.000019	$^{+	0.32	}_{-	0.31	}$	&	119.49	$^{+	9.19	}_{-	12.02	}$	&	2	\\
TOI-1730	&	01, 03	&	2:1	&	100.0	&	-0.14	$^{+	2.18	}_{-	1.92	}$	&	96.8	$^{+	5.43	}_{-	7.73	}$	&	2	\\
TOI-1746	&	01, 02	&	3:2	&	99.6	&	0.00059	$^{+	0.43	}_{-	0.41	}$	&	126.31	$^{+	10.11	}_{-	16.75	}$	&	2	\\
TOI-1749	&	b, c	&	2:1	&	100.0	&	0.15	$^{+	1.09	}_{-	1.31	}$	&	94.00	$^{+	7.76	}_{-	10.69	}$	&		\\
TOI-1803	&	02, 01	&	2:1	&	100.0	&	-0.037	$^{+	0.92	}_{-	0.87	}$	&	87.39	$^{+	13.63	}_{-	16.95	}$	&	2	\\
TOI-2086	&	02, 01	&	2:1	&	100.0	&	0.21	$^{+	2.12	}_{-	2.57	}$	&	84.69	$^{+	17.40	}_{-	18.11	}$	&	2	\\
TOI-2096	&	01, 02	&	2:1	&	100.0	&	0.00090	$^{+	0.30	}_{-	0.24	}$	&	87.38	$^{+	12.99	}_{-	17.05	}$	&	2	\\
TOI-2267	&	03, 01	&	3:2	&	100.0	&	0.0091	$^{+	0.39	}_{-	0.43	}$	&	122.83	$^{+	8.51	}_{-	12.67	}$	&	2	\\
TOI-4495	&	02, 01	&	2:1	&	100.0	&	0.0016	$^{+	0.66	}_{-	0.68	}$	&	100.39	$^{+	4.01	}_{-	5.26	}$	&	2	\\
\enddata
\tablecomments{For each new resonance, we include the system's name, the planets in resonance, the resonance, the percentage of simulations where this angle librates, the center of libration, and the amplitude of libration.\\ 1: System contains potential additional resonant pair (see Table~\ref{tab:nonconfirm}) \\ 2: Resonant pair contains candidate planet \\ 3: System has known TTVs \\ 4: Pair previously studied for resonance}
\end{deluxetable*}

\startlongtable
\begin{deluxetable*}{lccccccc}
\renewcommand{\arraystretch}{1.0}
\tablecolumns{8}
\tablewidth{\textwidth}
\tabletypesize{\tiny}
\tablecaption{Initial Conditions for Systems in Table~\ref{tab:confirm} \label{tab:conditions}}
\tablehead{
\colhead{System, No. Planets} &
\colhead{$M_{\star}~[M_{\odot}]$} &
\colhead{} &
\colhead{} &
\colhead{} &
\colhead{} &
\colhead{} &
\colhead{} \\
\colhead{Planets} & \colhead{} & \colhead{$R_p~[R_{\oplus}]$}	&	\colhead{$P~[d]$}	&	\colhead{$t_0~[d]$}	&	\colhead{$M_p~[M_{\oplus}]$}	&	\colhead{$i[\degree]$}	&	\colhead{Ref.}
}
\startdata
HD 28109, 3 (4)	&	$1.26^{+0.08}_{-0.08}$	&		&		&		&		&		&		\\
b	&		&	$2.2^{+0.1}_{-0.1}$	&	$22.89104^{+0.00035}_{-0.00036}$	&	$2458344.81772^{+0.00757}_{-0.00757}$	&	$18.5^{+9.1}_{-7.6}$	&	$87.725^{+0.023}_{-0.012}$	&	[10]	\\
c	&		&	$4.23^{+0.11}_{-0.11}$	&	$56.00819^{+0.00194}_{-0.00202}$	&	$2458377.80109^{+0.00724}_{-0.00733}$	&	$7.9^{+4.2}_{-3.0}$	&	$89.543^{+0.093}_{-0.086}$	&	[10]	\\
d	&		&	$3.25^{+0.11}_{-0.11}$	&	$84.25999^{+0.00744}_{-0.00662}$	&	$2458355.67324^{+0.00432}_{-0.00432}$	&	$5.7^{+2.7}_{-2.1}$	&	$89.682^{+0.093}_{-0.082}$	&	[10]	\\
02	&		&	$2.01^{+0.90}_{-0.09}$	&	$31.32312^{+0.00354}_{-0.00354}$	&	$2459044.82593^{+0.05923}_{-0.05923}$	&	\nodata	&	\nodata	&	\nodata	\\
	&		&		&		&		&		&		&		\\
HIP 41378, 5	&	$1.26^{+0.23}_{-0.16}$	&		&		&		&		&		&		\\
b	&		&	$2.90^{+0.44}_{-0.44}$	&	$15.572098^{+0.000018}_{-0.000019}$	&	$2457152.2818^{+0.0012}_{-0.0012}$	&	\nodata	&	$88.8^{+0.8}_{-1.4}$	&	[36], [3]	\\
c	&		&	$2.56^{+0.40}_{-0.40}$	&	$31.70648^{+0.00024}_{-0.00019}$	&	$2457163.1609^{+0.0023}_{-0.0027}$	&	\nodata	&	$87.5^{+2.2}_{-1.4}$	&	[36], [3]	\\
d	&		&	$3.96^{+0.59}_{-0.59}$	&	$156^{+163}_{-78}$	&	$2457166.2604^{+ 0.0017}_{-0.0017}$	&	\nodata	&	$89.930^{+0.025}_{-0.018}$	&	[36], [20], [3]	\\
e	&		&	$5.51^{+0.77}_{-0.77}$	&	$131^{+61}_{-36}$	&	$2457142.0194^{+0.0010}_{-0.0010}$	&	\nodata	&	$89.910^{+0.220}_{-0.045}$	&	[36], [3]	\\
f	&		&	$10.2^{+1.4}_{-1.4}$	&	$324^{+121}_{-126}$	&	$2457186.91423^{+0.00039}_{-0.00038}$	&	\nodata	&	$89.980^{+0.009}_{-0.006}$	&	[36], [20], [3]	\\
	&		&		&		&		&		&		&		\\
\ldots & \ldots & \ldots  & \ldots  & \ldots  & \ldots  & \ldots  & \ldots \\
	&		&		&		&		&		&		&		\\
TOI-406, 0 (2)	&	$0.38\pm0.02$	&		&		&		&		&		&		\\
02	&		&	$1.27\pm3.3$	&	$6.61491\pm0.00003$	&	$2458385.388\pm0.002$	&	\nodata	&	\nodata	&	\nodata	\\
01	&		&	$1.96\pm0.45$	&	$13.17573\pm0.00003$	&	$2458388.567\pm0.001$	&	\nodata	&	\nodata	&	\nodata	\\
\enddata
\tablerefs{%
 [1] = \citet{2021MNRAS.504.5327A};
 [3] = \citet{2019AJ....157..185B};
 [4] = \citet{2013Sci...340..587B};
 [5] = \citet{2011ApJS..197....7C};
 [6] = \citet{2016ApJS..226....7C};
 [7] = \citet{2021MNRAS.508..195D};
 [8] = \citet{2021A+A...653A..41D};
 [9] = \citet{2018MNRAS.480L...1D};
[10] = \citet{2022MNRAS.515.1328D};
[11] = \citet{2012ApJ...750..114F};
[12] = \citet{2012ApJ...750..113F};
[13] = \citet{2019RAA....19...41G};
[14] = \citet{hadden2017};
[15] = \citet{hadden2014};
[16] = \citet{2016ApJS..225....9H};
[17] = \citet{jontof2016};
[18] = \citet{2021AJ....161..246J};
[19] = \citet{2019ApJS..244...11K};
[20] = \citet{2021A+A...649A..26L};
[21] = \citet{2020AJ....159...57L};
[22] = \citet{2022AJ....163..101L};
[23] = \citet{2022A+A...666A.154L};
[24] = \citet{2014ApJS..210...20M};
[25] = \citet{2014ApJ...783...53M};
[26] = \citet{2018AJ....155..136M};
[27] = \citet{2016ApJ...822...86M};
[28] = \citet{2016MNRAS.463.1831N};
[29] = \citet{2019A+A...623A..41P};
[30] = \citet{2019A+A...623A..41P};
[31] = \citet{2014ApJ...784...45R};
[32] = \citet{2022AJ....163..151S};
[33] = \citet{2013MNRAS.428.1077S};
[34] = \citet{2022ApJ...926..120V};
[35] = \citet{2021MNRAS.507.2154V};
[36] = \citet{2016ApJ...827L..10V};
[37] = \citet{2014ApJ...783....4W};
[38] = \citet{2013ApJS..208...22X};
[39] = \citet{xie2014}.
}
\tablecomments{For each system we study for resonance, the number of
    planets (including planetary candidates) and the stellar mass
    $M_{\star}$ in solar masses. For each planet in these systems,
    the planet's radius $R_p$ in Earth radii, orbital period $P$ in
    days, the mid-transit time $t_0$ in BJD, the planet's mass $M_p$
    in Earth masses, the sky-plane inclination $i$ in degrees, and
    the reference for these values. }
\tablecomments{Table \ref{tab:conditions} is published in its
    entirety in the machine-readable format. A portion is shown here
    for guidance regarding its form and content.}
\end{deluxetable*}

\subsubsection{Systems to follow-up}\label{sec:followup}
As reported in Tables~\ref{tab:confirm} and \ref{tab:nonconfirm}, we study numerous systems whose resonant angles we cannot confirm are librating, but whose planets could very well be in a resonant chain. When we apply our method of resonant confirmation to the resonant chains in Kepler-80 and Kepler-223 (see Section~\ref{sec:known}), we find that the resonant angles only librate in $\sim$50-90\% of the simulations, but do not all reach our 90\% cut-off; we would therefore not be able to confirm such resonances with our method, and indeed we would not be able to confirm similar resonant chains and would instead see their resonant angles librate in $\sim$50+\% of our simulations. Following this logic, any systems that we explore where a) the period ratios between adjacent planets suggest the planets could be in a chain of resonances and b) our methods result in some high (but not 90+\%) percentage of \textit{N}-body simulations with librating resonant angles \textit{could} be resonant but we are not able to confirm that resonance in this work. Instead, the proximity to resonance is likely to lead to large gravitational perturbations that would be detectable in the system's RVs or transit times as deviations from Keplerian orbits. 

We list such systems below in Table~\ref{tab:follow}. For each system, we estimate the planet's TTVs by subtracting a linear least square fit from the transit times from each of the \textit{N}-body simulations. Such resonances could be confirmed through TTV fitting, transit+RV fitting, or photodynamic fitting \citep[e.g., ][]{mills2017,macdonald2022}, and therefore these systems deserve follow-up.

\begin{deluxetable*}{llllc}
\tablecolumns{5}
\tabletypesize{\footnotesize}
\tablecaption{ Systems to follow up\label{tab:follow}}
\tablehead{
\colhead{System} &
\colhead{$m_V$} &
\colhead{Planets} &
\colhead{Estimated TTVs (min) }&
\colhead{Prior Dynamics Study} 
}
\startdata
K2-243 & 10.971 & b--01, 01--c & 1--4, 2--15, 1--7 &  \\
Kepler-104$^\dagger$ & 14.266 & c--d & 1--3, 1--4, 0--1 & [1] \\
Kepler-105$^\dagger$ &  12.981 & b--c & 1--4, 1--15, 1--60 & \\
KOI-1358** &  15.477 & 02--03, 03--04 & 1--12, 5--22, 1--2, 1--5 & [1], [3], [4] \\ Kepler-79 & 14.036 & b--c, c--d, d--e & 1--20, 8--60, 6--24, 10--210 & [1], [2], [5] \\ Kepler-416 & 14.166 & b--c, 03--04 & 1--2, 1--4, 1--5, 1--6 & [1] \\ Kepler-122 & 14.403 & b--c, e--f & 1--3, 1--2, 2--10, 6--65 & [1] \\ Kepler-402& 13.270 & b--c, d--e & 1--8, 1--3, 2--9, 2--8, 1--6 & \\
Kepler-31$^\dagger$ & 15.496 & d--04 & 1--5, 3--85, 12--64, 2--77 & [1] \\ TOI-1136 &  9.534 & 02--01--04, 02--01, 01--04 & $\sim3800$, $\sim1300$, $\sim8400$, $\sim6000$ & \\
TOI-178$^\dagger$ &  11.955 & b--c--d, b--c, c--d, d--e & 1--14, 5--133, 16--825, 8--320 & \\
TOI-797 & 13.689 & 01--03, 03--02 & 1--1240, 1--6, 1--8 & \\
Kepler-154$^\dagger$ & 14.646 & b--c & 0, 1--10, 1--5, 5--20, 1--6, 1--3 & \\
Kepler-169$^\dagger$ &  14.424 & b--c & 0, 1--4, 1--6, 1--2, 0 & \\
Kepler-176$^\dagger$ & 14.767 &  b--e & 0, 2--30, 6--36, 2--74 & [1] \\ Kepler-62$^\dagger$ & 13.965 & b--c & 0, 1--9, 0, 1--6, 1--3 & \\
Kepler-224$^\dagger$ & 15.801 & b--c & 1--2, 1--3, 1--4, 2--5 & \\
Kepler-226$^\dagger$ & 15.563 & b--c, c--d & 2--29, 1--16, 1--14 & [1], [6] \\ Kepler-254$^\dagger$ & 16.012 & b--c & $\sim$1, 4--64, 2--64 & [1], [6]  \\ Kepler-374 &  14.701 & c--d, d--04 & 1--1300, 1--3, 1--3, 1--7, 1--2 & \\
Kepler-1518$^\dagger$** & 13.374 & 02--04 & 1--2, 1--2, 2--3  &  
\enddata
\tablecomments{Estimated TTV amplitudes, in minutes, for each planet in systems with possible resonant chains. {We also include each star's V (Johnson) magnitude as recorded in the exoplanet archive.} These planets are likely in a chain of resonances, but we are unable to confirm them with the methods we apply in this work. If the dynamics of the system has been studied before, we include a reference to the work. \\ $^\dagger$ System contains a confirmed resonance, either confirmed by this work or a previous work. \\ **Since our original study, the planets in KOI-1358 have been confirmed; the system is now Kepler-1987 \citep{valizadegan2023}. KOI-3741.04 has been confirmed as Kepler-1518c \citep{valizadegan2023}. \\ Ref: [1] \citet{hadden2014}, [2] \citet{jontof2014}, [3] \citet{jontof2016}, [4] \citet{hadden2017}, [5] \citet{yoffe2021}, [6] \citet{quinn2023}}
\end{deluxetable*}

\subsubsection{Shifted libration centers}

Without additional perturbations, a two-body resonance should always librate about 0$\degree$. We therefore identify systems below in Table~\ref{tab:shift180} whose angles librate about 180$\degree$ instead, suggesting that an additional resonant angle might be librating. These systems are interesting, but an in-depth study of their dynamics is beyond the scope of this work.

\begin{deluxetable}{lll}
\renewcommand{\arraystretch}{1.0}
\tablecolumns{3}
\tabletypesize{\normalsize}
\tablecaption{ Confirmed Resonant Angles Librating About 180$\degree$\label{tab:shift180}}
\tablehead{
\colhead{System} &
\colhead{Planets} &
\colhead{Additional Res.} 
}
\startdata
K2-239 & b, c & None, $\Theta_{c,d}$ at 6.6\% \\
TOI-406 & 02, 01 & None \\
Kepler-105 & c, 03 & None, $\Theta_{b,c}$ at 37\% \\
Kepler-18 & c, d & None, $\Theta_{b,c}$ at 13\% \\
Kepler-176 & d, e & $\Theta_{c,d}$ confirmed \\
Kepler-51 & b, c & $\Theta_{c,d}$ confirmed \\
Kepler-207 & c, d & $\Theta_{b,c}$ 22\% \\
Kepler-208 & c, d & None \\
Kepler-32 & b, c &  $\Theta_{e,b}$ confirmed \\
Kepler-1518 & b, 02 & None, $\Theta_{02,04}$ 79.4\%
\enddata
\tablecomments{For each confirmed angle that librates about 180$\degree$ instead of 0$\degree$, we list the system name, the planets involved in the resonance, and information about additional librating angles in the system. For additional angles that are not confirmed, we list the percentage of simulations quite resulted in libration.}
\end{deluxetable}

\subsection{Constraining the masses of Kepler-305}

Kepler-305 is a K-type star hosting three super-Earths and one mini-Neptune, with orbital periods ranging between 3.2 and 16.7~days. The orbital periods of the three outer planets suggest a chain of mean motion resonances of 1:2:3. The inner planet Kepler-305e sits just wide of the 5:3 resonance with Kepler-305b; although not a strong resonance, the larger mass of planet b \citep[$10.5^{+2.6}_{-2.0}~M_{\oplus}$,][]{xie2014} might be sufficient to lock the pair into resonance.

Kepler-305b and Kepler-305c exhibit anti-correlated TTVs which \citet{xie2014} used to confirm the planet pair and constrain their masses to $10.5^{+2.6}_{-2.0}~M_{\oplus}$ and $6.0^{+2.4}_{-2.2}~M_{\oplus}$, respectively. More recently, \citet{hadden2017} studied all four planets in the system, including the then-candidate Kepler-305e, fitting the system's TTVs to recover planetary masses and orbits. Despite robustly constraining the masses of both Kepler-305 c and Kepler-305 d and noting how close the three planets are to perfect commensurability, they do not find any of their fits to be resonant.

We model the Kepler-305 system via \textit{N}-body simulations using REBOUND. We initialize each planet with an orbital period, mid-transit time, eccentricity, and inclination drawn from independent, normal distributions centered on the nominal values from \citet{xie2014} and with widths equal to the uncertainties. We assume a stellar mass of 0.76$M_{\odot}$ \citep{xie2014}. We use a timestep of 5\% the inner planet's orbital period and integrate the system for 2~Myr using the WHFast integrator \citep{whfast}.

All of the 500 simulations we perform survive the 2~Myr integration without experiencing instability. We explore each simulation for resonance, looking for libration of the critical resonant angles 
$\phi_1 = 2\lambda_c -3\lambda_b + \lambda_e$, 
$\phi_2 = \lambda_d -2\lambda_c + \lambda_b$, 
$\Theta_{e,b} = 5\lambda_b - 3\lambda_e - 2\omega_b$, 
$\Theta_{b,c} = 3\lambda_c -2\lambda_b - \omega_c$, and 
$\Theta_{c,d} = 2\lambda_d - \lambda_c - \omega_c$. 
We summarize our results in Table~\ref{tab:kep305}.

\begin{deluxetable}{lccc}
\renewcommand{\arraystretch}{1.0}
\tablecolumns{4}
\tablewidth{\textwidth}
\tabletypesize{\normalsize}
\tablecaption{ Kepler-305 Resonances \label{tab:kep305}}
\tablehead{
\colhead{System} &
\colhead{\% librating} &
\colhead{Center} &
\colhead{Amplitude} 
}
\startdata
$\Theta_{e,b}$	&	0.0\%	& 	\nodata &	\nodata \\
$\Theta_{b,c}$	&	99.6\%	&	0.011	$^{+	1.35	}_{-	1.24	}$ &	89.8	$^{+	21.33	}_{-	24.5	}$ \\
$\Theta_{c,d}$	&	20.4\%	&-0.041	$_{-1.98}^{+2.24}$ & 138.71$_{-32.13}^{+7.37}$ \\
$\phi_1$	&	0.00\%	&	\nodata &	\nodata \\
$\phi_2$	&	1.6\%	&	\nodata &	\nodata 
\enddata
\tablecomments{Resulting three-body and two-body angles from our REBOUND \textit{N}-body simulations, including the libration centers and amplitudes in degrees. We find that the angle $\Theta_{b,c}$ is librating in nearly all of our simulations, so we confirm the two planets are resonant.}
\end{deluxetable}

Since the resonant angle $\Theta_{b,c}$ librates in 99.6\% of our \textit{N}-body simulations, we are able to conclude that planets b and c are in a 3:2 resonance. We use this confirmed resonance to constrain the masses of these two planets. We report the median of the distribution of masses in simulations with librating angles and the 16th and 84th percentile as the lower and upper uncertainties, respectively. In addition, we further estimate the masses of the other two planets, e and d, using the simulations with librating angles. We note that to use resonances to constrain planetary parameters, the resonances must be confirmed, so the mass estimates we report for planets e and d should not be seen as anything more than proof-of-concept. We summarize our mass estimates in Table~\ref{tab:K305_mass}.

\begin{deluxetable}{lcc}
\tablecolumns{3}
\tablewidth{2.0\columnwidth}
\tabletypesize{\normalsize}
\tablecaption{ Mass Estimates for Kepler-305 \label{tab:K305_mass}}
\tablehead{
\colhead{Planet} &
\colhead{Angle} &
\colhead{$M_p~[M_{\oplus}]$} 
}
\startdata
\hspace{7.5mm}e & $\phi_2*$ & 4.4$^{+0.22}_{ -0.21}$ \\
\hspace{7.0mm}b & $\Theta_{b,c}$ & 10.3$^{+2.7}_{ -2.3}$ \\
\hspace{7.0mm}b & $\phi_2*$ & 11.7$^{+2.5}_{ -1.5}$ \\
\hspace{7.0mm}c & $\Theta_{b,c}$ & 6.1$^{+2.6}_{ -2.5}$ \\
\hspace{7.0mm}c & $\Theta_{c,d}*$ & 5.8$^{+2.6}_{ -2.5}$ \\
\hspace{7.0mm}c & $\phi_2*$ & 6.6 $^{+2.2}_{ -1.4}$ \\
\hspace{6.9mm}d & $\Theta_{c,d}*$ & 8.6$^{+7.1}_{ -4.4}$
\enddata
\tablecomments{Mass estimates in $M_{\oplus}$ for each planet of Kepler-305. These estimates are the median with 16th and 84th percentile uncertainties of the distribution of mass for simulations where each angle is librating. \\ $*$ We are not able to confirm these resonances but include these mass estimates as proof-of-concept. }
\end{deluxetable}

We estimate the bulk density of each planet by drawing a radius estimate from a normal distribution centered on the nominal value from \citet{xie2014} with a width of the published uncertainties for each planet mass in our simulations with librating angles. We find that planets b and c likely have inflated atmospheres with bulk densities of $1.16_{-0.60}^{+ 1.72}~g/cm^3$ and $0.82_{ - 0.44}^{ + 1.44}~g/cm^3$, respectively, and that planet d could also be a mini-Neptune ($\rho=2.29_{- 1.27}^{+ 2.66}~g/cm^3$). Since planet e was only in resonance in 8 \textit{N}-body simulations, we do not estimate a density, although its size ($R_p=1.7^{+0.11}_{-0.08}~R_{\oplus}$) and proximity to its host star ($P=3.2$~days) suggest it is terrestrial.

\section{Known limitations}\label{sec:limitations}

\subsection{Cut-off for confirmed resonances}\label{sec:cutoff}
\begin{figure}
    \centering
    \includegraphics[width=0.9\columnwidth]{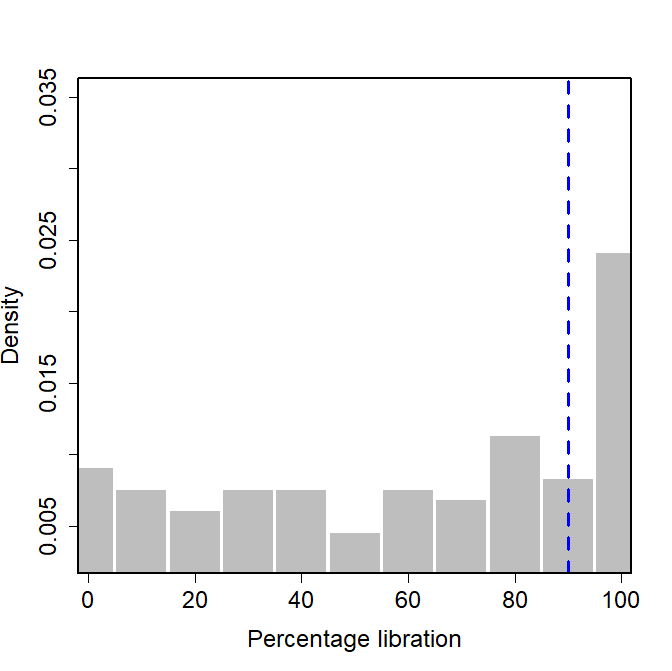}
    \caption{Distribution of percentages of simulations with librating angles. We note two modes, roughly separated by an absolute minimum at 53\%. We use the cutoff of 90\%, marked by the blue dashed line, to confirm a resonance.}
    \label{fig:percentages}
\end{figure}

In this work, we confirm a resonance if 90\% or more of the simulations result in librating angles. This cutoff of 90\% is arbitrary and intentionally high to avoid false positives. In Figure~\ref{fig:percentages}, we show the distribution of percentages of simulations with librating angles throughout all of the suites of simulations for this work. This distribution is roughly bimodal, 
with an absolute minimum at roughly 53\%. At a cut-off of 90\%, marked as a blue line in Figure~\ref{fig:percentages}, we are minimizing, but not eliminating, the number of false positives, but we are also likely failing to confirm truly resonant systems. We discuss these potential false negatives above in Section~\ref{sec:followup} and again stress the importance of additional, independent studies on these systems. 

\subsection{Measuring libration amplitudes}

\name~quantifies the libration amplitude as twice the angle's standard deviation (2$\sigma$). We explored variations of this measurement, including a more simplistic and traditional method of the difference between the minimum and maximum values over a period of time and the median absolute deviation, but these methods struggled with amplitudes of slowly circulating angles, identifying the angles as librating. Of these three methods, we found the 2$\sigma$ method to be better at marking truly librating angles and better at quantifying the amplitudes of angles that phase in and out of resonance; however, other works that have studied exoplanetary resonances use other methods, so a direct comparison of the libration amplitudes is not feasible. 

\newpage

\subsection{Confirming resonances using libration}
There are numerous other ways to study and confirm mean motion resonances, including verifying that the system lies within the separatrix of the system's phase space \citep{winter1997}. Our methods rely entirely on the libration of the resonant angles, and recent studies have suggested this technique might not be completely accurate; \citet{goldberg2022} found that many TTV systems might show librating angles, even if the system is formally nonresonant because the Hamiltonian has no separatrix.

In this work, we study 20 systems and 36 resonances that overlap with \citet{goldberg2022}. Of these systems, we confirm 40\% of resonances that they also confirm. In addition, we confirm 
nine resonances that \citet{goldberg2022} mark as not in resonance and fail to confirm twelve other resonances that they mark as resonant. Most of these twelve resonances are in multi-planet systems, with angles that librate in a large fraction of our simulations that we discuss in Section~\ref{sec:followup}.

{In addition, the two-body resonant angles that we study for libration might not fully describe the dynamics of the system. Depending on the apsidal angles of the two planets, a planet pair could be resonant with one or both resonant angles circulating \citep{laune2022}. Therefore, it is possible that some of the systems we were not able to confirm as resonant \textit{are} resonant, but their two-body angles circulate. We leave the expansion of \name~ to study the behaviour of $\Delta\bar{\omega}=\bar{\omega_1}-\bar{\omega_2}$ and the behaviour of the mixed resonant angle defined in Eq. 39 of \citet{laune2022} to future work. }

\subsection{Inflated libration amplitudes}

The libration amplitudes of the resonant angles can provide information into a system's formation history and subsequent evolution; they are to some degree reliant on the eccentricities of the planets when they lock into resonance \citep{mustill2011}, and the amplitudes can also grant insight into the system's stability and rate of migration \citep[e.g., ][]{goldreich2014, hadden2020}.  Ideally, then, we would be able to study each of these newly confirmed resonances and place constraints on their stability and formation history, but unfortunately, the amplitudes we recover through this method appear artificially inflated. In fact, the inflation of libration amplitudes was first noted by \citet{millholland2018} and later explored by \citet{jensen2022}; they found that noisy data can lead to libration amplitudes that are systematically biased towards larger values.

We test to see if this bias could be artificially inflating our libration amplitudes. We perform two additional suites of \textit{N}-body simulations of Kepler-363. In the first suite, we reduce the uncertainties on the planet masses from 30\% to 10\%, and in the second suite, we inflate the uncertainties of the orbital elements by 100\% their measured values. In total, we explore three situations: 1) average mass uncertainty and average orbital uncertainty, 2) small mass uncertainty and average orbital uncertainty, and 3) average mass uncertainty and large orbital uncertainty. We then study these systems as we have the real systems, using \name~to characterize the resonances. We find that we recover inflated amplitudes for each of the three the cases with well constrained masses and orbital parameters as with more poorly constrained parameters; a Kolmogorov–Smirnov
two-sample test between the recovered libration amplitudes from each suite of simulations results in large \textit{p}-values ($p>0.05$), failing to reject the null hypothesis that the samples are drawn from the same population. Because of this, we are not able to claim that the uncertainties in our planetary masses or orbital parameters are driving factors behind our inflated amplitudes. 

We do, however, recover some interesting additional results that might provide insight:
\begin{itemize}
    \item  Suite 2 (reduced mass uncertainties) has a narrower eccentricity distribution than Suites 1 and~3. \item The libration amplitude of $\Theta_{b,c}$ is larger with smaller $e_b$ (negatively correlated).
    \item When mass uncertainties are reduced (Suite 2), we see a positive correlation between the libration amplitude of $\Theta_{b,c}$ and $e_b$.
    \item The libration amplitude of $\Theta_{c,d}$ is larger with smaller $e_c$ (negatively correlated) and larger $e_d$ (positively correlated).
    \item Correlation strength between libration amplitudes and eccentricities depends on how eccentricities compare to one another. Lower mass uncertainties result in these correlations becoming much stronger.
    \begin{itemize}
    \item If $e_c<e_d$, then the libration amplitude of $\Theta_{b,c}$ is greater with larger $e_b$ and $e_c$ (positively correlated), and the libration amplitude of $\Theta_{c,d}$ is greater with larger $e_c$ and $e_d$ (positively correlated).
    \item If $e_b>e_c$, then the libration amplitude of $\Theta_{b,c}$ is greater with larger $e_b$ and $e_c$ (positively correlated), but the libration amplitude of $\Theta_{c,d}$ is greater with \textit{smaller} $e_c$ and $e_d$ (negatively correlated).
 \end{itemize}
\end{itemize}

\subsection{{Exoplanet Archive}}
{As described above in Section~\ref{sec:findnew}, we pull the inputs to our analysis from the \texttt{exoplanet archive}, and therefore the results we present in Section~\ref{sec:findnew} are dependent on the parameters being correct. The parameters reported as default parameters on the \texttt{exoplanet archive} span a range of quality and precision and result from various methods of measurement and estimation. Although we attempt to mitigate this variety by exploring a large number of simulations with parameters drawn from independent distributions, there likely still exist underlying biases that skew our results. By comparing the output of \name~ with dynamical integrations of RV or TTV fits, we can vet our results and determine how reliant they are on the accuracy and precision of the input parameters. We leave such a study to future work.}

{In addition to the planetary parameters, we pull the stellar mass from the \texttt{exoplanet archive} and keep a fixed mass for all simulations. Although the host mass does not directly impact the resonant nature of a system, no study\footnote{\citet{matsumoto2020} find that \textit{existing} resonant chains can be broken if the host loses mass.} has explored how much of an impact, if any, a different host mass could have on the resonant state or libration centers or amplitudes of librating resonant angles. We leave such a study to future work.}

\subsection{Resonant Chains}

As noted above in Section~\ref{sec:known}, \name~struggles to confirm resonant chains and resonances in systems with more than four planets. Because of this, we are not able to confirm many resonant chains that would likely greatly improve our numbers. 

We try to mitigate this limitation with our discussion in Section~\ref{sec:followup} and Table~\ref{tab:follow}, as these systems will require additional follow-up as \name~cannot be the sole method of study.

\subsection{Expensive}
The methods employed in this work are very computationally intensive. They are expensive and only possible at this large scale on a high-performance computing cluster. Ideally, we would individually characterize enough two-body resonances and resonant chains to perform more informed searches, making use of various machine learning techniques to improve performance. 

\section{Conclusion}\label{sec:conclusion}

The study of mean motion resonances (MMRs) allows for the unique constraint of planetary formation and evolution as well as the constraint of the planets' masses and orbital parameters. Although MMRs are so information-rich, it can be challenging and computationally intensive to confirm two planets are actually in resonance with one another. Because of these factors, we have had relatively few confirmed resonant systems.

Following the methods of \citet{macdonald2022} and \citet{quinn2023}, we create the python package \name~\citep{exommr} to identify, confirm, and study new mean motion resonances in the exoplanetary population. Our methods rely on suites of \textit{N}-body simulations in rebound, and we confirm a resonance if 90\% or more of the simulations result in librating resonant angles. We recover the known resonances in Kepler-80 and Kepler-223, noting the shortcomings of this method for resonant chains. We demonstrate this software's capabilities by constraining orbital parameters and masses of planets in known resonances and by constraining the parameter space of unknown large planets in well-studied resonant systems.

After verifying the software's abilities and demonstrating its use, we search the Kepler/K2 and TESS catalogues for new resonances. We identify 66 new resonant systems, including seven new resonant chains. We describe the limitations of the software, and include a list of resonances we were not able to confirm but are likely to librate. These systems deserve follow-up analysis, either through different methods or with additional data, since they are likely to demonstrate detectable TTV signals. 

Our methods herein are computationally intensive and infeasible on personal machines with small numbers of CPUs. We intend to confirm additional resonances until we have a large enough population for more informed and AI-trained searches, but such an analysis is beyond the scope of this work. {In addition, our methods for confirming resonance are limited as described in Section~\ref{sec:limitations}, potentially leading to a high false negative rate and missing resonances. We will explore how our methods of resonant confirmation (e.g., the use of librating angles, amplitude estimate) ultimately affect our ability to confirm resonance and compare our results to RV and TTV fitting in follow-up studies. }

{We thank the anonymous referee for the constructive feedback that improved this manuscript. }The authors acknowledge use of the ELSA high performance computing cluster at The College of New Jersey for conducting the research reported in this paper. This cluster is funded in part by the National Science Foundation under grant numbers OAC-1826915 and OAC-1828163. This research has made use of the NASA Exoplanet Archive, which is operated by the California Institute of Technology, under contract with the National Aeronautics and Space Administration under the Exoplanet Exploration Program.

\bibliographystyle{aasjournal}
\bibliography{citations.bib}

\appendix

Below, in Table~\ref{tab:nonconfirm}, we include the resonances we explored but were not able to confirm. We include systems with planetary candidates and note potential resonances that are in systems with confirmed resonances; we have found that resonant chains are challenging to confirm with this method, so it is possible these systems have more than the singular confirmed resonance. In addition, there exist various reasons why \name~ could fail to confirm a known resonance. We will explore these reasons, outlined in Section~\ref{sec:limitations}, in follow-up studies.

\startlongtable
\begin{deluxetable}{lcccc}
\renewcommand{\arraystretch}{1.0}
\tablecolumns{5}
\tablewidth{\textwidth}
\tabletypesize{\small}
\tablecaption{ Other potential resonances \label{tab:nonconfirm}}
\tablehead{
\colhead{System} &
\colhead{Planets} &
\colhead{Resonance} &
\colhead{\% librating} &
\colhead{Notes} 
}
\startdata
K2-19	&	d, b	&	3:2	&	0.00	&		\\
K2-19	&	b, c	&	3:2	&	67.21	&		\\
K2-37	&	b, c	&	4:3	&	0.00	&		\\
K2-37	&	c, d	&	2:1	&	37.40	&		\\
K2-72	&	b, d	&	4:3	&	0.00	&	1	\\
K2-72	&	d, c	&	2:1	&	0.00	&	1	\\
K2-72	&	c, e	&	3:2	&	0.00	&	1	\\
K2-178	&	04, 05	&	4:3	&	0.00	&	1, 2	\\
K2-239	&	c, d	&	3:2	&	6.60	&	1	\\
K2-243	&	b, 01	&	3:2	&	58.60	&	2	\\
K2-243	&	01, c	&	4:3	&	44.20	&	2	\\
K2-266	&	c, d	&	2:1	&	0.00	&		\\
K2-266	&	d, e	&	4:3	&	32.68	&		\\
K2-268	&	b, d	&	2:1	&	26.00	&	1	\\
K2-268	&	d, e	&	4:3	&	9.20	&	1	\\
K2-285	&	c, d	&	4:3	&	0.00	&	1	\\
K2-285	&	d, e	&	4:3	&	0.00	&	1	\\
K2-384	&	c, d	&	3:2	&	5.65	&		\\
K2-384	&	d, e	&	4:3	&	0.00	&		\\
K2-384	&	e, f	&	4:3	&	21.47	&		\\
Kepler-18	&	b, c	&	2:1	&	13.00	&	1	\\
Kepler-23	&	c, d	&	4:3	&	0.00	&	1	\\
Kepler-31	&	b, c	&	2:1	&	5.40	&	1	\\
Kepler-31	&	d, 04	&	2:1	&	70.60	&	1, 2	\\
Kepler-32	&	c, d	&	5:2	&	0.0	&		\\
Kepler-33	&	d, e	&	4:3	&	0.0	&		\\
Kepler-33	&	e, f	&	5:4	&	1.2	&		\\
Kepler-53	&	d, b	&	2:1	&	36.8	&	1	\\
Kepler-62	&	b, c	&	2:1	&	40.4	&	1	\\
Kepler-62	&	c, d	&	4:3	&	0.0	&	1	\\
Kepler-79	&	b, c	&	2:1	&	59.20	&		\\
Kepler-79	&	c, d	&	2:1	&	79.20	&		\\
Kepler-79	&	d, e	&	3:2	&	77.40	&		\\
Kepler-83	&	d, b	&	2:1	&	25.2	&	1	\\
Kepler-84	&	d, b	&	2:1	&	63.8	&		\\
Kepler-84	&	b, c	&	4:3	&	0.0	&		\\
Kepler-84	&	c, e	&	2:1	&	0.6	&		\\
Kepler-102	&	b, c	&	4:3	&	19.80	&	1	\\
Kepler-102	&	c, d	&	4:3	&	0.00	&	1	\\
Kepler-102	&	e,f	&	5:3	&	0.00	&	1	\\
Kepler-104	&	c, d	&	2:1	&	60.80	&	1	\\
Kepler-105	&	b, c	&	3:2	&	37.20	&	1	\\
Kepler-114	&	b, c	&	3:2	&	3.60	&		\\
Kepler-114	&	c, d	&	4:3	&	0.00	&		\\
Kepler-122	&	b, c	&	2:1	&	83.40	&		\\
Kepler-122	&	e, f	&	3:2	&	82.00	&		\\
Kepler-131	&	b, c	&	5:3	&	28.20	&	1	\\
Kepler-138	&	c, d	&	5:3	&	0.00	&	1	\\
Kepler-154	&	e, f	&	5:2	&	0.00	&	1	\\
Kepler-154	&	b, c	&	2:1	&	67.00	&	1	\\
Kepler-169	&	b, c	&	2:1	&	50.80	&	1	\\
Kepler-176	&	d, e	&	2:1	&	89.00	&	1	\\
Kepler-184	&	b, c	&	2:1	&	75.80	&		\\
Kepler-184	&	c, d	&	4:3	&	0.40	&		\\
Kepler-207	&	b, c	&	2:1	&	22.40	&	1	\\
Kepler-208	&	d, e	&	4:3	&	0.0	&	1	\\
Kepler-215	&	b, c	&	3:2	&	18.0	&		\\
Kepler-215	&	c, d	&	2:1	&	82.4	&		\\
Kepler-217	&	d, b	&	4:3	&	33.80	&		\\
Kepler-217	&	b, c	&	3:2	&	22.60	&		\\
Kepler-224	&	b, c	&	2:1	&	72.0	&	1	\\
Kepler-226	&	b, c	&	4:3	&	42.0	&		\\
Kepler-226	&	c, d	&	3:2	&	45.8	&		\\
Kepler-249	&	b, c	&	2:1	&	82.8	&	1	\\
Kepler-254	&	b, c	&	2:1	&	42.4	&	1	\\
Kepler-271	&	d,c	&	4:3	&	0.00	&		\\
Kepler-271	&	c, d	&	4:3	&	17.80	&		\\
Kepler-271	&	b, 04	&	5:3	&	0.00	&	2	\\
Kepler-271	&	04, 05	&	5:4	&	0.00	&	2	\\
Kepler-292	&	b, c	&	4:3	&	0.0	&		\\
Kepler-292	&	c, d	&	2:1	&	86.2	&		\\
Kepler-292	&	d, e	&	5:3	&	5.2	&		\\
Kepler-305	&	e, b	&	5:3	&	0.00	&	1	\\
Kepler-305	&	c, d	&	2:1	&	20.40	&	1	\\
Kepler-326	&	b, c	&	2:1	&	67.0	&		\\
Kepler-326	&	c, d	&	4:3	&	0.0	&		\\
Kepler-327	&	c, d	&	5:2	&	0.0	&		\\
Kepler-339	&	b, c	&	4:3	&	3.00	&	1	\\
Kepler-350	&	b, c	&	3:2	&	10.8	&		\\
Kepler-350	&	c, d	&	4:3	&	0.0	&		\\
Kepler-352	&	04, d	&	4:3	&	0.00	&	2	\\
Kepler-352	&	d, b	&	4:3	&	0.00	&		\\
Kepler-363	&	c, d	&	3:2	&	68.4	&	1	\\
Kepler-374	&	b, c	&	5:3	&	0.6	&		\\
Kepler-374	&	c, d	&	3:2	&	79.6	&		\\
Kepler-374	&	d, 04	&	3:2	&	74.2	&	2	\\
Kepler-374	&	04, 05	&	3:2	&	0.0	&	2	\\
Kepler-394	&	d, b	&	4:3	&	0.0	&		\\
Kepler-402	&	b, c	&	3:2	&	69.80	&		\\
Kepler-402	&	c, d	&	4:3	&	0.00	&		\\
Kepler-402	&	d, e	&	5:4	&	87.60	&		\\
Kepler-402	&	e, 05	&	4:3	&	0.00	&	2	\\
Kepler-416	&	b, c	&	2:1	&	45.80	&		\\
Kepler-416	&	c, 03	&	2:1	&	0.00	&	2	\\
Kepler-416	&	03, 04	&	2:1	&	82.80	&	2	\\
Kepler-431	&	b, c	&	5:4	&	25.20	&		\\
Kepler-431	&	c, d	&	4:3	&	40.20	&		\\
Kepler-968	&	b, c	&	3:2	&	13.80	&	1	\\
Kepler-1073	&	c, 04	&	3:2	&	59.40	&	2	\\
Kepler-1073	&	04, b	&	4:3	&	1.20	&	2	\\
Kepler-1130	&	04, c	&	3:2	&	0.00	&	2	\\
Kepler-1130	&	c, d	&	4:3	&	0.00	&		\\
Kepler-1130	&	d, b	&	5:4	&	30.80	&		\\
Kepler-1371	&	c, b	&	4:3	&	0.00	&		\\
Kepler-1371	&	03, 04	&	5:4	&	0.00	&	2	\\
Kepler-1371	&	04, 05	&	5:4	&	0.66	&	2	\\
Kepler-1518	&	02, 04	&	2:1	&	79.4	&	1, 2	\\
Kepler-1542	&	c, b	&	4:3	&	8.60	&		\\
Kepler-1542	&	b, e	&	5:4	&	0.20	&		\\
Kepler-1542	&	d, 05	&	5:4	&	0.00	&	2	\\
Kepler-1581	&	b, 02	&	4:3	&	0.00	&	1, 2	\\
Kepler-1693	&	c, 04	&	3:2	&	0.00	&	2	\\
Kepler-1693	&	04, b	&	3:2	&	45.60	&	2	\\
Kepler-1693	&	b, 03	&	3:2	&	0.00	&	2	\\
KOI-1358	&	01, 02	&	3:2	&	6.20	&	2	\\
KOI-1358	&	02, 03	&	3:2	&	48.00	&	2	\\
KOI-1358	&	03, 04	&	3:2	&	75.60	&	2	\\
KOI-3083	&	01, 02	&	4:3	&	0.00	&	2	\\
KOI-3083	&	02, 03	&	5:4	&	0.00	&	2	\\
TOI-178	&	b, c, d	&	1:2:3	&	63.20	&	1	\\
TOI-178	&	b, c	&	2:1	&	68.00	&	1	\\
TOI-178	&	c, d	&	3:2	&	76.40	&	1	\\
TOI-178	&	d, e	&	2:1	&	79.20	&	1	\\
TOI-270	&	b, c	&	5:3	&	0.40	&	1	\\
TOI-421	&	b, c	&	3:1	&	0.20	&		\\
TOI-700	&	04, d	&	4:3	&	85.40	&	2	\\
TOI-797	&	01, 03	&	3:2	&	60.80	&	2	\\
TOI-797	&	03, 02	&	3:2	&	50.00	&	2	\\
TOI-1136	&	02, 01, 04	&	1:2:3	&	26.56	&	2	\\
TOI-1136	&	02, 01	&	2:1	&	75.89	&	2	\\
TOI-1136	&	01, 04	&	3:2	&	43.97	&	2	\\
TOI-1136	&	04, 03	&	4:3	&	0.00	&	2	\\
TOI-1246	&	b, c	&	4:3	&	31.40	&	1	\\
TOI-1246	&	c, d	&	3:1	&	0.00	&	1	\\
\enddata
\tablecomments{Each additional potential resonance explored, including system name, planet pair, resonance explored, and percentage of simulations with librating angle. \\ 1: System contains confirmed resonant pair \\ 2: Pair contains at least one candidate planet}
\end{deluxetable}

\newpage

\end{document}